\documentclass[12]{article}
\usepackage{fleqn}
\usepackage{graphicx}
\usepackage{amsmath}
\begin{document}
\Large
\begin{center}
{\bf Charting the Real Four-Qubit Pauli Group via Ovoids of a Hyperbolic Quadric of PG(7,\,2)}
\end{center}
\vspace*{.5cm}
\large
\begin{center}
Metod Saniga,$^{1}$ P\' eter L\' evay$^{2}$ and Petr Pracna$^{3}$
\end{center}
\vspace*{-.3cm} \normalsize
\begin{center}
$^{1}$Astronomical Institute, Slovak Academy of Sciences,
SK-05960 Tatransk\' a Lomnica\\ Slovak Republic\\
(msaniga@astro.sk)\vspace*{.4cm}

$^{2}$Department of Theoretical Physics, Institute of Physics,
Budapest University of Technology\\ Budafoki \' ut. 8, H-1521
Budapest, Hungary\\(levay@neumann.phy.bme.hu)

\vspace*{.1cm}
and

\vspace*{.1cm}
$^{3}$J. Heyrovsk\' y Institute of Physical
Chemistry, v.v.i., Academy of Sciences of Czech Republic\\ Dolej\v skova 3, CZ-182 23 Prague 8, Czech Republic\\
(pracna@jh-inst.cas.cz)

\end{center}

\vspace*{.2cm} \noindent \hrulefill

\vspace*{.2cm} \noindent {\bf Abstract}

\noindent The geometry of the real four-qubit Pauli group, being embodied in the structure of the symplectic polar space $W(7,2)$, is analyzed in terms of ovoids of a hyperbolic quadric of PG$(7,2)$, the seven-dimensional projective space of order two. The quadric is selected in such a way that it contains all 135 symmetric elements of the group. Under such circumstances, the third element on the line defined by any two points of an ovoid is skew-symmetric, as is the nucleus of the conic defined by any three points of an ovoid. Each ovoid thus yields 36/84 elements of the former/latter type, accounting for all 120 skew-symmetric elements of the group. There are a number of notable types of ovoid-associated subgeometries of the group, of which we mention the following: a subset of 12 skew-symmetric elements lying on four mutually skew lines that span the whole ambient space, a subset of 15 symmetric elements that corresponds to two ovoids sharing three points, a subset of 19 symmetric elements generated by two ovoids on a common point, a subset of 27 symmetric elements that can be partitioned into three ovoids in two unique ways, a subset of 27 skew-symmetric elements that exhibits a $15 + 2 \times 6$ split reminding that exhibited by an elliptic quadric of PG$(5, 2)$, and a subset of seven skew-symmetric elements formed by the nuclei of seven conics having two points in common, which is an analogue of a Conwell heptad of PG$(5, 2)$.

The strategy  we employed is completely novel and unique in its nature, mainly due to the fact that hyperbolic quadrics in binary projective spaces of higher dimensions have no ovoids. Such a detailed dissection of the geometry of the group in question may, for example, be crucial in getting further insights into the still-puzzling black-hole-qubit correspondence/analogy.

\hspace*{1.5cm}Hyperbolic Quadric of PG$(7, 2)$ -- Conwell Heptad

\vspace*{-.0cm} \noindent \hrulefill

\vspace*{.0cm}
\large

\section{Introduction}
It is already a firmly-established fact that the structure of the generalized Pauli group acting on the Hilbert space of $N$-qubits, $N \geq 2$, is embodied in the geometry of the symplectic polar space of rank $N$ and order two, $W(2N - 1, 2)$ \cite{sp}-\cite{pla}. The elements of the group (discarding the identity) answer to the points of $W(2N - 1, 2)$, their partitions into maximally commuting subsets correspond to spreads of the space, a maximally commuting subset has its representative in a maximal totally isotropic subspace of $W(2N - 1, 2)$ and, finally, commuting translates into collinear (or, perpendicular). In the case of the {\it real} $N$-qubit Pauli group, the structure of the corresponding symplectic polar space can be refined in terms of the orthogonal polar space $Q^{+}(2N - 1, 2)$ --- that is, a hyperbolic quadric of the ambient projective space PG$(2N - 1, 2)$ --- which is the locus of symmetric elements of the group \cite{hos}.

Due to great importance of this group in quantum information theory, the $N = 2$ case was analyzed in very detail \cite{ps,spp}. Here, the layout of nine symmetric elements of the group lying on the quadric $Q^{+}(3, 2)$ was found to be isomorphic to a copy of the so-called Mermin magic square --- an array furnishing one of the simplest proofs of the famous Kochen-Specker theorem \cite{ks}.
The geometry of the next, $N=3$ case turned out to be of great relevance for the so-called black-hole-qubit correspondence \cite{lsv}. Here, the quadric $Q^{+}(5, 2)$ can be found as an extension of the Levi graph of the Fano plane and occurs also as a geometric hyperplane of the split Cayley hexagon of order two \cite{lsv,vl}. Moreover, its 35 points are in a well-known bijection with 35 lines of PG$(3,2)$; this property not only provides a remarkable link between the two- and three-qubit Pauli groups, but also offers a nice finite-geometrical background for another elegant proof of the Kochen-Specker theorem \cite{sl}.

In this paper, we aim at dissecting the geometry of the (real) {\it four}-qubit Pauli group to the extent that can be compared with the previous two cases. The quadric $Q^{+}(7, 2)$ exhibits a high degree of symmetry, being unusual in that it admits a graph automorphism of order three, known as triality, that swaps its points and two systems of generators, and preserves the set of totally singular lines (see, e.\,g., \cite{study,tits}). The structure of this quadric can be visualized quite well through one of its ovoids, that is, a set of nine points which has exactly one point in common with every generator (maximal totally singular subspace). $Q^{+}(7, 2)$ was treated this way in exhaustive detail by Edge \cite{edge1}, which will be the standard reference for us and where the interested reader can look for more details.\footnote{It is worth mentioning here that the notation employed by Edge is nowadays rather obsolete; thus, for example, in his language `ovoid' reads `ennead' and for quadrics he uses the terms `ruled/non-ruled' instead of `hyperbolic/elliptic,' respectively. }

A number of distinguished geometrical sub-configurations of the ambient space PG$(7, 2)$ associated with a particular ovoid of $Q^{+}(7, 2)$ will be diagrammatically recast in terms of subsets of elements of the real four-qubit Pauli group, where a particular attention will be paid to those featuring both symmetric and skew-symmetric elements. Some of these subsets have already shown up in quantum physics, like the generalized quadrangle GQ$(2, 4)$; this remarkable finite geometry, which admits parametrization  by elements of both three-qubit and two-qutrit Pauli groups, underlies the geometry of the $E_{6}$-symmetric entropy formula of a certain class of stringy black holes \cite{lsvp} and, as we shall see, arises in our current setting from two ovoids sharing a triple of points. But  a majority of them are new, like an analogue of a Conwell heptad of PG$(5, 2)$, formed by the nuclei of seven conics on two common points of an ovoid.
Treating all sets/configurations in a unified, ovoid-based manner also makes this paper a good repository of noteworthy geometrical subsets of the real four-qubit Pauli group for any scholar dealing with generalized Pauli groups, whether of their own or in view of possible applications.

\section{Symplectic Polar Spaces, Quadrics, Ovoids and Conwell Heptads}
In this section we shall collect some basic, well-known facts about the geometrical concepts that will be employed in the sequel.

A (finite-dimensional) classical polar space (see, for example, \cite{ht,cam}) describes the geometry of a $d$-dimensional vector space
over the Galois field GF$(q)$, $V(d, q)$, carrying a non-degenerate reflexive
sesquilinear form $\sigma(x, y)$. The polar space is called symplectic,
and usually denoted as $W(d-1,q)$,  if this form is bilinear
and alternating, i.e., if $\sigma(x, x) = 0$ for all $x \in V(d,
q)$; such a space exists only if $d=2N$, where $N \geq 2$ is called its
rank. A subspace of $V(d, q)$ is called totally isotropic if
$\sigma$ vanishes identically on it. $W(2N-1,q)$ can then be
regarded as the space of totally isotropic subspaces of the ambient space PG$(2N-1,
q)$, the ordinary $(2N - 1)$-dimensional projective space over
$GF(q)$, with respect to a symplectic form (also known as a null
polarity), with its maximal totally isotropic subspaces, also
called {\it generators}, having dimension $N - 1$.  For $q=2$ this
polar space contains $|{\rm PG}(2N-1, 2)| = 2^{2N} - 1 = 4^{N} - 1$
points and $(2+1)(2^2+1)\cdots(2^N+1)$ generators.

A quadric in PG$(d, q)$, $d \geq 1$, is the set of points whose coordinates satisfy an equation of the form $\sum_{i,j=1}^{d+1} a_{ij} x_i x_j = 0$, where at least one $a_{ij} \neq 0$.
Up to transformations of coordinates, there is one or two distinct kinds of non-singular quadrics in PG$(d, q)$ according as $d$ is even or odd, namely \cite{ht}:
\begin{itemize}
\item $Q(2N,q)$, the {\it parabolic} quadric formed by all points of PG$(2N, q)$ satisfying the standard equation $x_1x_2+\cdots+x_{2N-1}x_{2N} + x_{2N+1}^{2} = 0$;
\item $Q^{-}(2N - 1,q)$, the {\it elliptic} quadric formed by all points of PG$(2N - 1, q)$ satisfying the standard equation $f(x_1,x_2)+x_3x_4+\cdots+x_{2N-1}x_{2N} = 0$, where $f$ is irreducible over GF$(q)$; and
\item $Q^{+}(2N - 1,q)$, the {\it hyperbolic} quadric formed by all points of PG$(2N - 1, q)$ satisfying the standard equation $x_1x_2+x_3x_4+\cdots+x_{2N-1}x_{2N} = 0$;
    \end{itemize}
where $N \geq 1$. A parabolic quadric is specific in that all its tangent hyperplanes have a common point --- called the {\it nucleus} (or kernel). As in the case of polar spaces, any subspace of maximal dimension that lies fully on a quadric is called its generator; the corresponding dimension is equal to $N - 1$ for parabolic and hyperbolic quadrics and $N - 2$ for elliptic ones.
The set of generators of  $Q^{+}(2N - 1,q)$, $N \geq 2$, is divided into two equally-sized disjoint families; two generators belong to the same family iff the co-dimension of their intersection has the same parity as $N - 1$.
The number of points and/or generators lying on quadrics is as follows \cite{ht}:
 \begin{itemize}
\item $|Q(2N,q)|_p = (q^{2N}-1)/(q-1)$,
\item $|Q^{-}(2N - 1,q)|_p = (q^{N-1}-1)(q^{N}+1)/(q-1) $,
\item $|Q^{+}(2N - 1,q)|_p = (q^{N-1}+1)(q^{N}-1)/(q-1)$,
    \end{itemize}
and/or
\begin{itemize}
\item $|Q(2N,q)|_g = (q+1)(q^2+1)\cdots(q^N+1)$,
\item $|Q^{-}(2N - 1,q)|_g = (q^2+1)(q^3+1)\cdots(q^{N}+1) $,
\item $|Q^{+}(2N - 1,q)|_g = 2(q+1)(q^2+1)\cdots(q^{N-1}+1)$,
    \end{itemize}
respectively.
Thus, for example, $Q^{+}(7, 2)$ features $(2^3+1)(2^4-1) = 135$ points and $2 \times 135 = 270$ generators.
An {\it ovoid} of a non-singular quadric is a set of points that has exactly one point common with each of its generators. An ovoid of $Q^{-}(2s - 1,q)$, $Q(2s,q)$ or $Q^{+}(2s + 1,q)$ has $q^s+1$ points; an ovoid of $Q^{+}(7, 2)$ comprises $2^3+1 = 9$ points.

Given the hyperbolic quadric $Q^{+}(2N - 1, q)$ of PG$(2N - 1, q)$, $N \geq 2$, a set $X$ of points such that each line joining two distinct points of $X$ has no point in common with $Q^{+}(2N - 1, q)$  is called an exterior set of the quadric. It is known  that $|X| \leq (q^N - 1)/(q - 1)$; if $|X| = (q^N - 1)/(q - 1)$, then $X$ is called a maximal exterior set. Maximal exterior sets are rather scarce and have already been completely classified \cite{thas}. It was found that $Q^{+}(5, 2)$  (the Klein quadric) has, up to isomorphism, a {\it unique} one --- also known, after its discoverer, as a {\it Conwell hetpad} \cite{con}. The set of 28 points lying off $Q^{+}(5, 2)$ comprises eight such heptads, any two having exactly one point in common.

\section{Generalized Pauli Groups and Symplectic Polar Spaces}
The generalized real $N$-qubit Pauli groups (see, e.\,g., \cite{nc}), ${\cal P}_N$, are generated by $N$-fold tensor products of the matrices
\begin{eqnarray*}
I = \left(
\begin{array}{cc}
1 & 0 \\
0 & 1 \\
\end{array}
\right),~
X = \left(
\begin{array}{cc}
0 & 1 \\
1 & 0 \\
\end{array}
\right),~
Y = \left(
\begin{array}{cc}
0 & -1 \\
1 & 0 \\
\end{array}
\right)
~{\rm and}~
Z = \left(
\begin{array}{cc}
1 & 0 \\
0 & -1 \\
\end{array}
\right).
\end{eqnarray*}
Explicitly,
\begin{equation*}
{\cal P}_N = \{\pm A_1 \otimes A_2 \otimes\cdots\otimes A_N:~ A_i \in \{I, X, Y, Z \},~ i = 1, 2,\cdots,N \}.
\end{equation*}
These groups are well known in physics and play an important role in the theory of quantum error-correcting codes (see, e.\,g., \cite{cald}), with $X$ and $Z$ being, respectively, a bit flip and phase error of a single qubit. Here, we are more interested in their factor groups $\overline{{\cal P}}_N \equiv {\cal P}_N/{\cal Z}({\cal P}_N)$, where the center ${\cal Z}({\cal P}_N)$ consists of $\pm I_{(1)} \otimes I_{(2)} \otimes \cdots \otimes I_{(N)}$. For a particular value of $N$, the $4^N - 1$ elements of $\overline{{\cal P}}_N \backslash \{I_{(1)} \otimes I_{(2)} \otimes \cdots \otimes I_{(N)}\}$  can be bijectively identified with the same number of points of
$W(2N-1, 2)$ in such a way that two commuting elements of the group will lie on the same totally isotropic line of this polar space \cite{sp}-\cite{th}; moreover, those elements of the group
whose square is $ + I_{(1)} \otimes I_{(2)} \otimes \cdots \otimes I_{(N)}$ (i.\,e., symmetric elements) are then found to lie on a certain  $Q^{+}(2N - 1,2)$ of the ambient space PG$(2N-1, 2)$ \cite{hos}. For the sake of completeness, it should also be added that generators, of both $W(2N - 1, 2)$ and $Q^{+}(2N - 1, 2)$, correspond to {\it maximal} sets of mutually commuting elements of the group.

As already mentioned in the introduction, the symplectic geometry of the {\it two}-qubit Pauli group, $W(3, 2)$, is also isomorphic to the smallest non-trivial generalized quadrangle, GQ$(2, 2)$, and the corresponding hyperbolic quadric, $Q^{+}(3, 2)$, the locus of nine symmetric elements of the group, is the geometry behind the Mermin(-Peres) magic square \cite{ps,spp}. Maximal sets of five pairwise non-commuting elements of the group have geometrical counterparts in ovoids of GQ$(2, 2)$, or, equivalently, elliptic quadrics  $Q^{-}(3, 2)$ of the ambient space PG$(3, 2)$ \cite{hos}. The {\it three}-qubit Pauli group is notable by the fact that its 35 symmetric elements, being located on a Klein quadric $Q^{+}(5, 2)$ of PG$(5, 2)$, can be identified via the Klein correspondence with 35 lines of PG$(3,2)$, and, via a particular copy of $W(3, 2)$ in the latter space, with certain three-elements sets of the two-qubit Pauli group; moreover, the 28 skew-symmetric elements of the group located outside the Klein quadric form eight distinct copies of Conwell heptads. Maximal sets of pairwise commuting operators are of cardinality seven and correspond to Fano planes in the associated $W(5, 2)$.

\section{Dissecting the Four-Qubit Case via Ovoids of  $Q^{+}(7, 2)$}
\subsection{Preliminaries}

We now come to the core section, the one devoted to revealing a plethora of fine traits of the $W(7, 2)$-geometry of the factored {\it four}-qubit Pauli group.

We shall start with elementary observation that 255 distinct elements of this group split into 135 symmetric (i.\,e., those squaring to $+I \otimes I \otimes I \otimes I$) and 120 skew-symmetric  (i.\,e., those squaring to $-I \otimes I \otimes I \otimes I$). Any maximal set of mutually commuting elements features 15 elements, being geometrically isomorphic to a generator of $W(7, 2)$; altogether, there are   $(2+1)(2^2+1)(2^3+1)(2^4+1) = 2295$ such sets (see Sec.\,2). In what follows we shall be more concerned with the set of 135 symmetric operators and, hence, a particular copy of $Q^{+}(7, 2)$ which they all are found to lie on.

Our first task is to set up a bijective mapping between the elements of the group and the points of the associated polar/ambient projective space.
If we take a basis of $W(7,2)$ in which the sesquilinear form $\sigma(x,y)$ is given by
\begin{eqnarray*}
\sigma(x,y) =
(x_1 y_5 - x_5 y_1) + (x_2 y_6 - x_6 y_2) + (x_3 y_7 - x_7 y_3) + (x_4 y_8 - x_8 y_4),
\end{eqnarray*}
then this bijection acquires the form:
\begin{equation}
A_i \leftrightarrow (x_i, x_{i+4}),~i \in \{1, 2, 3, 4\},
\end{equation}
with the understanding that
\begin{equation}
I \leftrightarrow (0,0),~X \leftrightarrow (0,1),~Y \leftrightarrow (1,1),~
Z \leftrightarrow (1,0);
\end{equation}
thus, for example, the point having coordinates $(0,1,1,0,0,1,0,1)$ corresponds to the element $I \otimes Y \otimes Z \otimes X$. It then follows that the equation of the $Q^{+}(7, 2)$ accommodating all symmetric elements must have the following standard form
\begin{equation}
x_1x_5 + x_2x_6 + x_3x_7 + x_4x_8 = 0.
\end{equation}
This can readily be inspected using the fact that the matrix $Y$ is the only skew-symmetric element in the set $\{I, X, Y, Z\}$ (see Sec.\,3) and, so, any symmetric element of the group must contain an even number of $Y$s. In his seminal paper \cite{edge1}, to be substantially drawn on as mentioned already in introduction, Edge works with a different system of homogeneous coordinates, namely with the one where the equation of $Q^{+}(7, 2)$ features all 28 distinct products of pairs:
\begin{equation}
\sum_{i < j} y_iy_j = 0.
\end{equation}
We shall, therefore, be in need of the transformation relating the  two coordinate systems, viz.
\begin{eqnarray}
&&x_1 = y_1 + y_4 + y_6 + y_8, \nonumber \\
&&x_2 = y_2 + y_3 + y_6 + y_8, \nonumber\\
&&x_3 = y_2 + y_4 + y_5 + y_8, \nonumber\\
&&x_4 = y_2 + y_4 + y_6 + y_7, \nonumber\\
&&x_5 = y_3 + y_5 + y_8, \nonumber\\
&&x_6 = y_4 + y_7 + y_8, \nonumber\\
&&x_7 = y_2 + y_3 + y_7, \nonumber\\
&&x_8 = y_1 + y_2 + y_8.
\end{eqnarray}
One of the reasons Edge had for such selection was that nine points of one of the ovoids lying on  $Q^{+}(7, 2)$ have a particularly simple and appealing coordinate expression
\begin{eqnarray}
&&(1,0,0,0,0,0,0,0), \nonumber \\
&&(0,1,0,0,0,0,0,0), \nonumber \\
&&(0,0,1,0,0,0,0,0), \nonumber \\
&&(0,0,0,1,0,0,0,0), \nonumber \\
&&(0,0,0,0,1,0,0,0), \nonumber \\
&&(0,0,0,0,0,1,0,0), \nonumber \\
&&(0,0,0,0,0,0,1,0), \nonumber \\
&&(0,0,0,0,0,0,0,1), \nonumber \\
&&(1,1,1,1,1,1,1,1),
\end{eqnarray}
which in our coordinates acquires a more complex form
\begin{eqnarray}
&&(1,0,0,0,0,0,0,1) \leftrightarrow Z \otimes I \otimes I \otimes X, \nonumber \\
&&(0,1,1,1,0,0,1,1) \leftrightarrow I \otimes Z \otimes Y \otimes Y, \nonumber \\
&&(0,1,0,0,1,0,1,0) \leftrightarrow X \otimes Z \otimes X \otimes I, \nonumber \\
&&(1,0,1,1,0,1,0,0) \leftrightarrow Z \otimes X \otimes Z \otimes Z, \nonumber \\
&&(0,0,1,0,1,0,0,0) \leftrightarrow X \otimes I \otimes Z \otimes I,  \nonumber \\
&&(1,1,0,1,0,0,0,0) \leftrightarrow Z \otimes Z \otimes I \otimes Z,  \nonumber \\
&&(0,0,0,1,0,1,1,0) \leftrightarrow I \otimes X \otimes X \otimes Z,  \nonumber \\
&&(1,1,1,0,1,1,0,1) \leftrightarrow Y \otimes Y \otimes Z \otimes X,  \nonumber \\
&&(0,0,0,0,1,1,1,1) \leftrightarrow X \otimes X \otimes X \otimes X.
\end{eqnarray}
$Q^{+}(7, 2)$ contains 960 ovoids in total, forming a single orbit under its automorphism group \cite{edge1}. Hence, without loss of generality, any of them can be taken to work with and we shall opt for the above given one.

By its very definition (see Sec.\,2), an ovoid ${\cal O}$ of the $Q^{+}(7, 2)$ is a set of nine points that has exactly one point in common with every PG$(3, 2)$ lying fully on the $Q^{+}(7, 2)$.\footnote{The attentive reader may ask whether there exists a simple `prescription' for picking up an ovoid out of the 135 points of the quadric, or if its is possible to give its algebraic characterization as eq. (3) plus some extra constraint(s), instead of going through a rather lengthy chain of calculations implied by the original `combinatorial' definition. As far as we know, the answer to both the questions is negative.}  Moreover, any $s$-point subset of  ${\cal O}$, $2 \leq s \leq 7$, defines a unique PG$(s-1,2)$ of the ambient space PG$(7, 2)$; otherwise stated, no $k$ points of ${\cal O}$, $3 \leq k \leq 8$, lie in the same PG$(k-2, 2)$. It is these subspaces and their intersections with the $Q^{+}(7, 2)$ that will be, after being reinterpreted by eqs. (1) and (2) in terms of the subsets of the associated four-qubit Pauli group, of most interest to us. To find these structures, we shall employ an elementary fact that the coordinates of the third point on the line defined by two distinct points of PG$(7, 2)$ can be found as the sum of the coordinates of the two points; for example, the third point on the line defined by the points $(0,1,1,1,0,1,0,0)$ and $(1,1,0,0,0,1,1,0)$ is (note that $1+1=0$ in GF(2))
\begin{equation*}
(0,1,1,1,0,1,0,0)+ (1,1,0,0,0,1,1,0) = (1,0,1,1,0,0,1,0).
\end{equation*}
Under our bijection, this {\it sum} is seen to transform into ordinary (matrix) {\it product} of the corresponding group elements,
\begin{equation*}
(I \otimes Y \otimes Z \otimes Z)(Z \otimes Y \otimes X \otimes I) = Z \otimes  I \otimes Y \otimes Z.
\end{equation*}
This remarkable `sum-goes-into-product' rule, together with bijection (1) and (2), will enable us a quick and straightforward transition/move between the finite-geometrical and group-theoretical settings. Our analysis will be accompanied by a number of illustrations/pictures, which should help the reader  grasp the essentials of the concepts dealt with and guide his visualization of them.

\begin{figure}[t]
\vspace*{-1.5cm}
\centerline{\includegraphics[width=14cm,clip=]{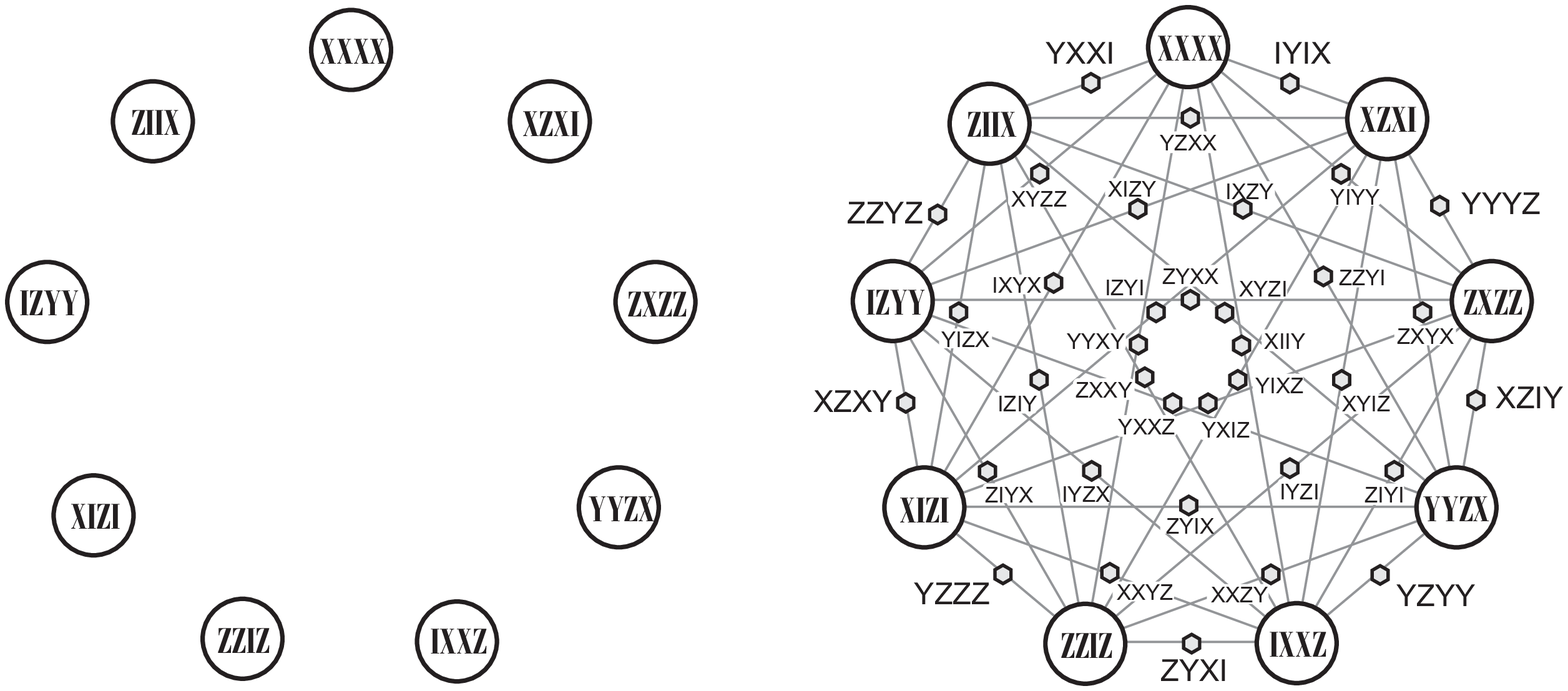}}
\vspace*{-1.5cm}
\caption{{\it Left}: A diagrammatical illustration of the ovoid ${\cal O^{*}}$. Its nine points are represented by circles and labelled by the nine corresponding elements of the associated Pauli group. To save space, in this and all subsequent figures $ A_1 A_2 A_3 A_4$ is shorthand for $A_1 \otimes A_2 \otimes A_3 \otimes A_4$. {\it Right}: The set of 36 skew-symmetric elements of the group that corresponds to the set of third points of the lines defined by pairs of points of our ovoid. The lines are illustrated by dashes and the elements in question by small shaded hexagons. This pictorial distinction between symmetric (circles) and skew-symmetric (hexagons) elements will also be used in the sequel.}
\end{figure}

\begin{figure}[pth!]
\centerline{\includegraphics[width=5.5cm,clip=]{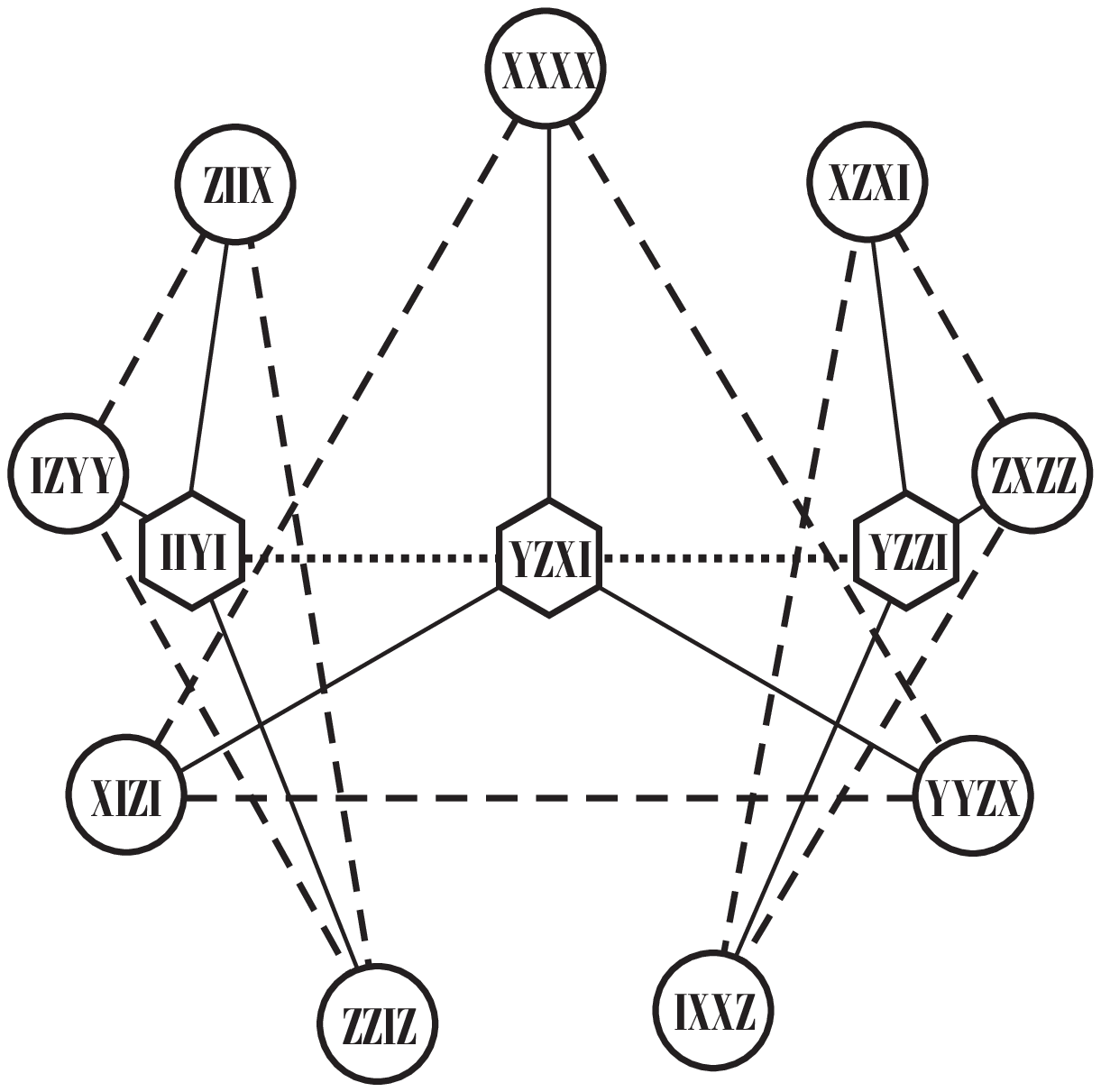}\hfill\includegraphics[width=5.5cm,clip=]{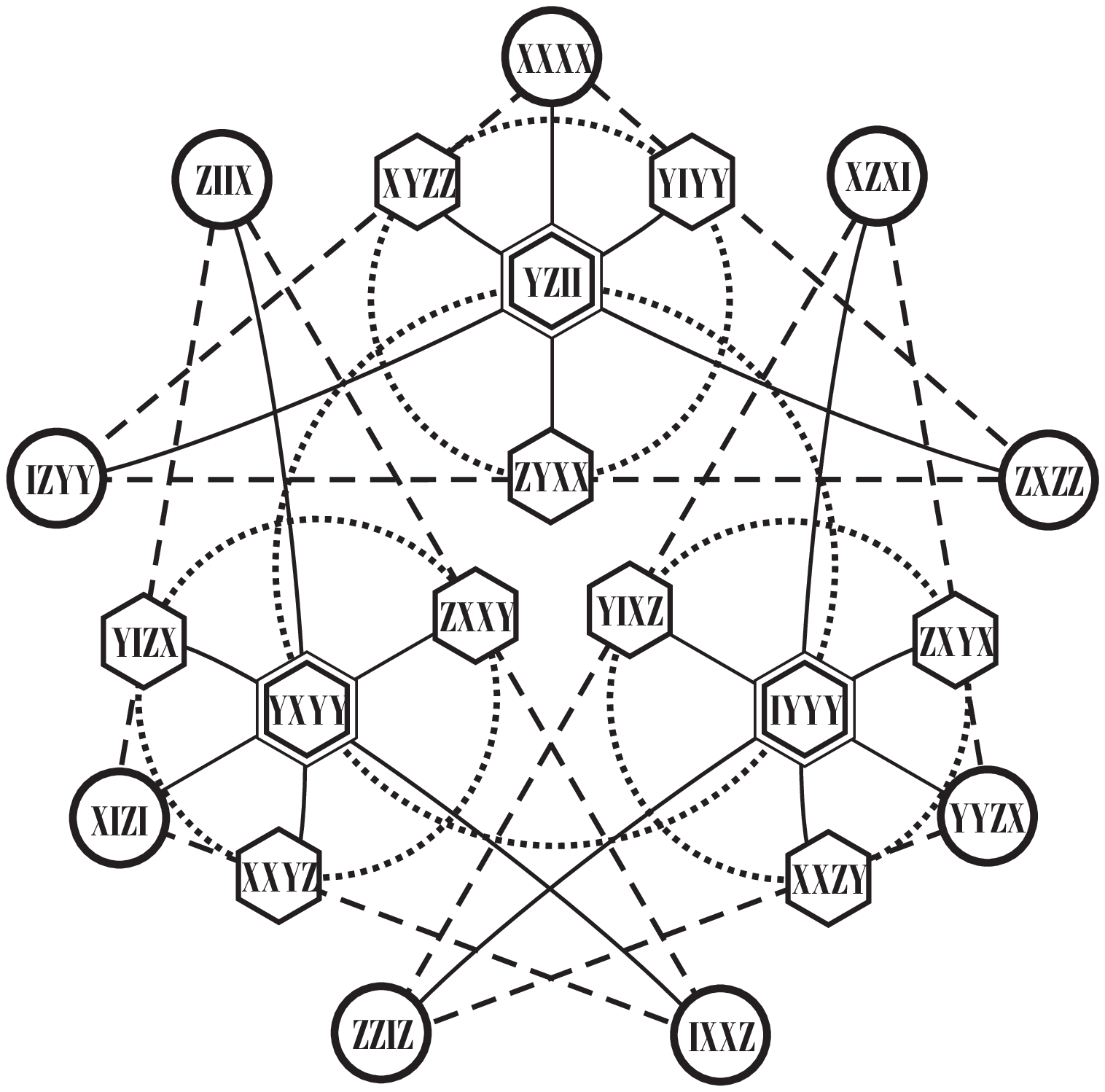}}
\vspace*{.2cm}
\caption{{\it Left}: A partition of our ovoid into three conics (vertices of dashed triangles) and the corresponding axis (dotted). As the coordinates of the nucleus of a conic are the sum of those of its three points, the group element associated with the nucleus is the product of the three elements of the conic.  {\it Right}: The tetrad of mutually skew, off-quadric lines (dotted) characterizing a particular partition of ${\cal O^{*}}$; also shown in full are the three Fano planes associated with the partition.}
\end{figure}

\subsection{Subconfigurations of the Group Related to Intersections of\\ Projective Subspaces with the $Q^{+}(7, 2)$}

Our point of departure is an ovoid ${\cal O}$ of the $Q^{+}(7, 2)$, in particular examples the one defined by eq.\,(7), denoted by ${\cal O^{*}}$ and also sketched in Figure 1, {\it left}. We shall first look at the lines (i.\,e., PG$(1, 1)$s) defined by pairs of its points. As the third point of such a line lies off the $Q^{+}(7, 2)$ \cite{edge1}, the corresponding element of the group is skew-symmetric. There are obviously ${9 \choose 2}  = 36$ such secant lines and because any two of them are disjoint/skew for otherwise four points of the ovoid would be in the same plane, a contradiction, so there are as well 36 different skew-symmetric elements --- as depicted in Figure 1, {\it right}.

Next we pass to subspaces defined by triples of points of ${\cal O}$, i.\,e. to Fano planes (PG$(2, 2)$s). Any such triple represents a conic (i.\,e., a parabolic quadric $Q(2, 2)$) and
is the only intersection of  $Q^{+}(7, 2)$ with the associated Fano plane \cite{edge1}. The nucleus of the conic is thus an off-quadric point and answers to a
skew-symmetric element. Any ovoid generates ${9 \choose 3} = 84$ distinct conics; since no two of them share their nuclei, this is also the number of different nuclei. From the properties
of ${\cal O}$ it further follows that this set of 84 skew-symmetric elements is disjoint from the 36-element one defined above and so the union of the two sets accounts for all 120 skew-symmetric
elements of $\overline{{\cal P}}_4$. A remarkable fact is that if ${\cal O}$ is partitioned into three sets of three points, the nuclei of the corresponding conics lie on a line --- clearly a line
skew to $Q^{+}(7, 2)$ \cite{edge1}. Such a line is called an axis of  ${\cal O}$ and for a particular partition of our selected ovoid ${\cal O^{*}}$ it is shown in Figure 2, {\it left}. Obviously,
every ovoid yields $ {9 \choose 3}{6 \choose 3}/3!  = 280 $ partitions and, so, the same number of axes. As any Fano plane contains just one line skew to the quadric (namely the polar of its nucleus),
any such partition defines a unique {\it tetrad} of lines that are pairwise disjoint and lying all off the quadric \cite{edge1}; namely, the three lines originating from the Fano planes and
the axis --- see Figure 2, {\it right}. The four lines of a tetrad are notable in that they span the whole PG$(7, 2)$.  Sets of 12 skew-symmetric elements that correspond to such tetrads of lines,
totalling to 11200 \cite{edge1}, are thus another kind of distinguished subsets of $\overline{{\cal P}}_4$.
\begin{figure}[t]
\centerline{\includegraphics[width=5.5cm,clip=]{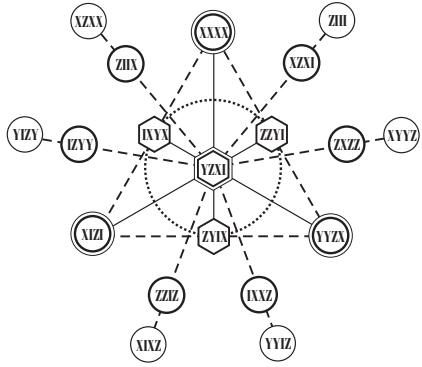}}
\vspace*{.2cm}
\caption{A conic (doubled circles) of ${\cal O^{*}}$ (thick circles), is located in another ovoid (thin circles). The six lines through the nucleus of the conic (dashes) pair the distinct points of the two ovoids. Also shown is the ambient Fano plane of the conic.}
\end{figure}
Each triad of points of an ovoid ${\cal O}$ is shared by one more ovoid, ${\cal O'}$. The 12 points of the symmetric difference of the two can be grouped into six pairs, each having points from both ovoids, in such a way that the six lines they define all pass through a common point --- the nucleus of the conic defined by the three shared points; an illustration of this
is remarkable `double-six' property is given in Figure 3.
\begin{figure}[h]
\centerline{\includegraphics[width=7.5cm,clip=]{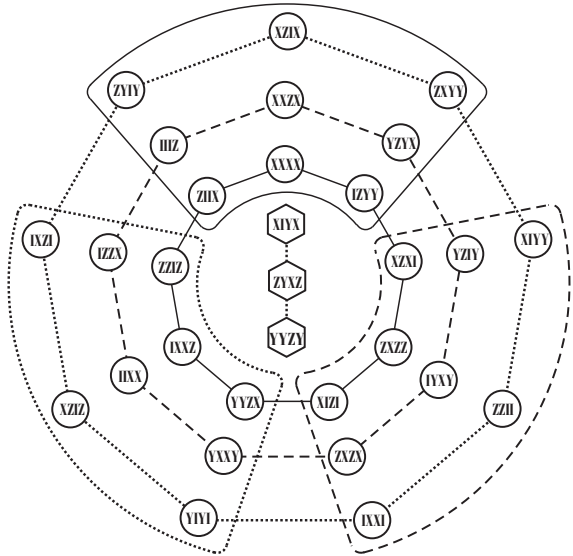}}
\vspace*{.2cm}
\caption{An example of the set of 27 symmetric operators of the group that can be partitioned into three ovoids in two distinct ways. The six ovoids, including ${\cal O^{*}}$ (solid nonagon), have a common axis (shown in the center).}
\end{figure}

This property has a very interesting implication, namely  that any partition of an ovoid into three triples of points leads to a {\it unique} set of three mutually disjoint ovoids, each of which shares with the original ovoid one triple. And, even more interestingly, the set of 27 points comprising these three ovoids can be split into three mutually disjoint ovoids in another way, where one of the ovoids is that we started with.  So, one arrives at  a remarkable set of six ovoids falling into complementary triads, where each member of one triad shares three points with each member of the other, and all of them are on the same axis. Edge \cite{edge1}, on page 19,  gives an explicit example of such configuration. This  configuration, after being transformed by (5) into our system of coordinates and then mapped into the group-setting by (1) and (2), looks as illustrated in Figure 4.
Here, ovoids of one triad are represented by vertices of three concentric nonagons  (solid, dashed and dotted)  and those of the other triad by vertices located in three non-overlapping areas  delineated again by solid, dashed and dotted segments. From the group-theoretical point of view it is worth noting that any symmetric element commutes with five elements of any of the six ovoids (see Figure 5, {\it left}), whereas a skew-symmetric element commutes either with three or seven elements in each of the six ovoids, or with three elements in four ovoids and seven elements in two ovoids (see Figure 5, {\it right}).

\begin{figure}[t]
\centerline{\includegraphics[width=7.2cm,clip=]{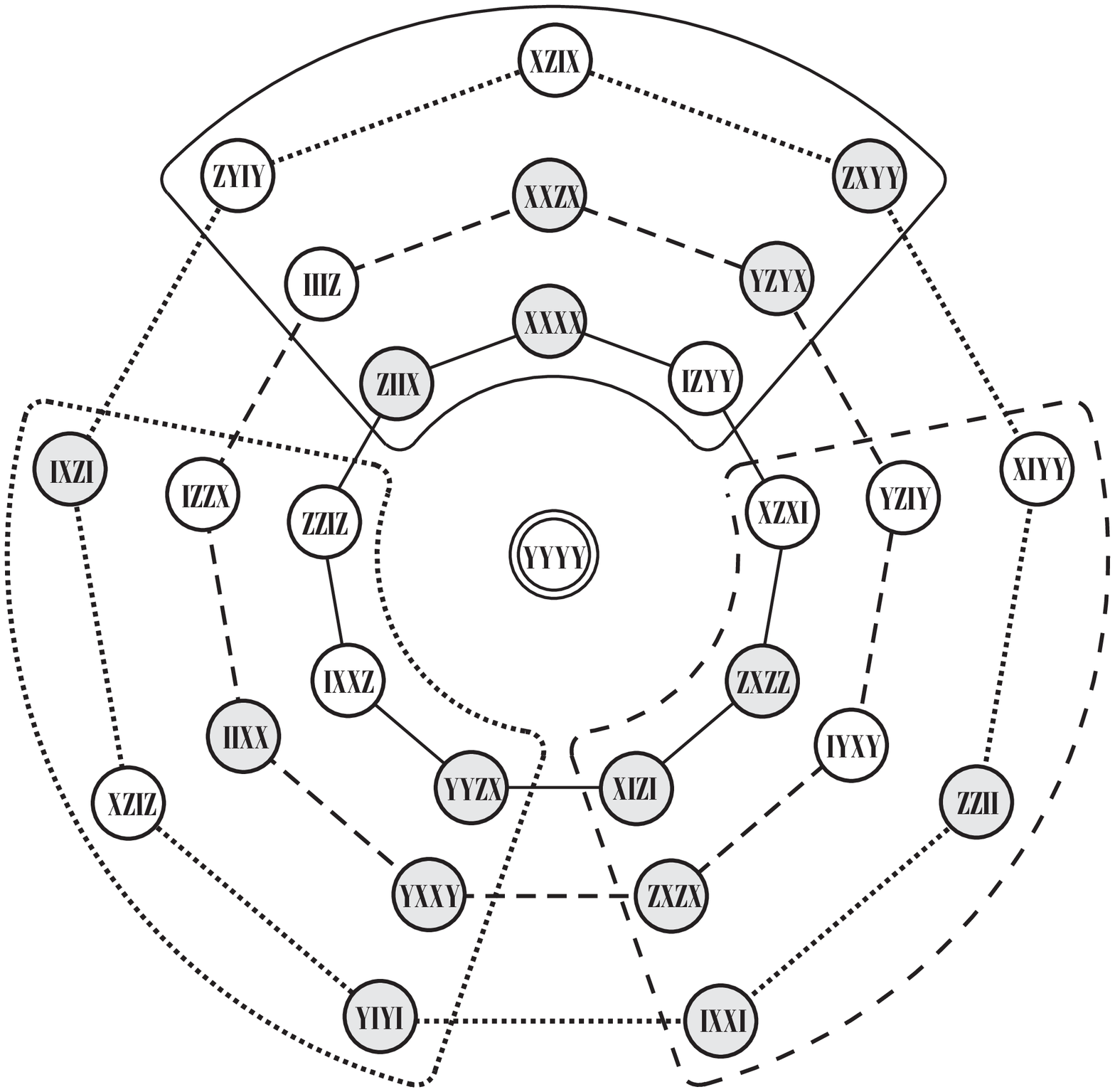}\hfill\includegraphics[width=7.2cm,clip=]{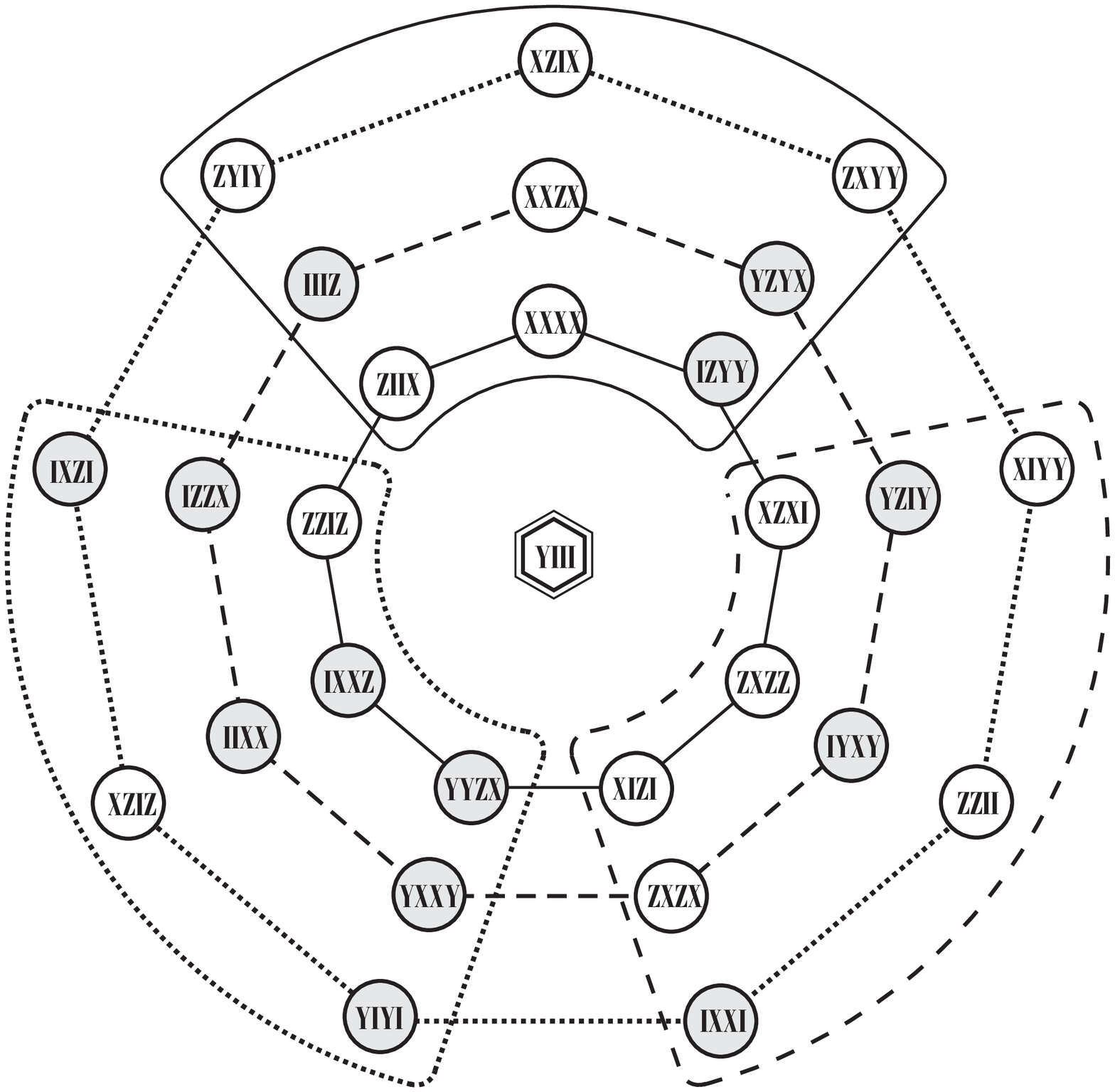}}
\vspace*{.2cm}
\caption{The commuting/non-commuting property of the 27 elements  of the previous figure with respect to a symmetric ({\it left}) and a skew-symmetric ({\it right}) element of the group. In both cases, shaded are those elements that commute with the element shown in the center. }
\end{figure}

 Our next move is  to subspaces PG$(3, 2)$s, henceforth called solids, which are  defined by quadruples of points of ${\cal O}$. Any such solid shares with the $Q^{+}(7, 2)$ one more point  that, clearly, does not belong to ${\cal O}$, the five points lying on a $Q^{-}(3, 2)$ of the solid in question. As there are ${9 \choose 4} = 126$ solids associated with a given ${\cal O}$, these additional points, which are all distinct, account for all the remaining 126 points of the $Q^{+}(7, 2)$ \cite{edge1}. If one selects a point of ${\cal O}$ and partition the remaining eight points into two quadruples, the two additional/supplementary points of the two solids defined by the quadruples are found to lie on a line that passes through the selected point \cite{edge1}; given ${8 \choose 4}/2= 35$
 such partitions, there are 35 such lines through the point in question.  Let us have a more detailed look at one such partition of our particular ${\cal O}^{*}$, as portrayed in Figure 6. The point selected is represented by a double-circle and a partition, together with the corresponding additional points, by shaded and non-shaded circles. The four lines joining the additional point of one solid and the four points of the second solid lie fully on the $Q^{+}(7, 2)$. The set of eight more points  we get this way (see Figure 6) and the point selected form an ovoid as well. As each partition of ${\cal O}^{*}$ yields a different ovoid, from what we said earlier it follows that on a given point of an ovoid there are 35 other ovoids having with it no other point in common \cite{edge1}.
 One can, of course, also reverse the preceding chain of reasoning. That is, one starts with a pair of ovoids having a single point in common and find the unique line through the common point and lying fully on the $Q^{+}(7, 2)$ whose two additional points lead to the required complementary partition of either of the two ovoids.
 It is instructive to compare this case with that of two ovoids on a common conic (Figure 3).

 \begin{figure}[t]
\centerline{\includegraphics[width=6.5cm,clip=]{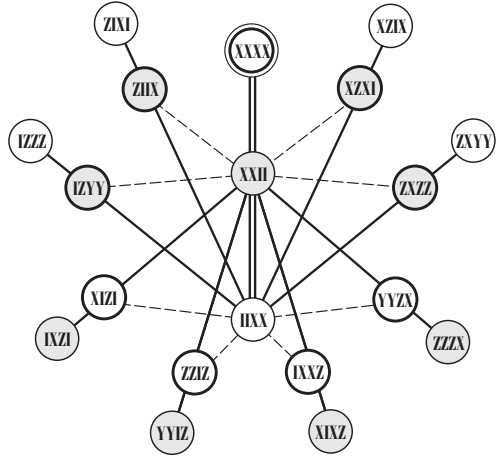}}
\vspace*{.2cm}
\caption{Two ovoids, one being ${\cal O}^{*}$, sharing a point ($X \otimes X \otimes X \otimes X$), and a particular $1+4+4$ partition of both that shows how they are related to each other through two additional points ($X \otimes X \otimes I \otimes I$ and  $I \otimes I \otimes X \otimes X$) of the solids defined by the partition. Disregarding the common point, the remaining 18 points (symmetric group elements) of the configuration can also be viewed as two pairs of $Q^{-}(3, 2)$, each `concurring' on one of the two additional points.}
\end{figure}

At this point, we shall make a slight digression from our main line to draw the reader's attention to the following intriguing facts. It is known \cite{edge1} that there are 64 ovoids passing through any point of  $Q^{+}(7, 2)$. Choosing one of them as a reference, there will be 63 sharing with it this particular point. We already know that 35 of them are such that each has no other point in common with the chosen ovoid. And since two non-disjoint ovoids can share either one or three points \cite{edge1}, each of the remaining 28 ovoids must be of the second type; and this is indeed the case as a point of an ovoid is the meet of ${8 \choose 2} = 28$ triangles. This split reminds us of a similar split of 63 points of  PG$(5, 2)$ into 35/28 points lying on/off a Klein quadric $Q^{+}(5, 2)$. As PG$(5, 2)$ is the ambient space of the symplectic polar space $W(5, 2)$ that underlies the real three-qubit Pauli group, mapping the above-described set of 63 ovoids into the set of points of PG$(5, 2)$ provides us with a  very interesting relation between {\it single} elements of the three-qubit group and specific {\it nine}-element sets of the four-qubit group --- a relation that also accounts for the fundamental $35 + 28$ split.\footnote{Another important relation between the two groups is furnished by the bijection that sends 135 generators  of $W(5, 2)$ into 135 points of the orthogonal polar space $Q^{+}(7, 2)$ of $W(7, 2)$, that is a mapping where a {\it maximum set} of mutually commuting elements of the three-qubit group corresponds to a {\it single} symmetric element of the four-qubit group.}

Let us return back from our detour and proceed to subspaces that are generated by pentads of points of $Q^{+}(7, 2)$, i.\,e. to PG$(4, 2)$s. As any pentad contains ${5 \choose 4} = 5$ quartets, the corresponding PG$(4, 2)$ comprises the five associated solids, each contributing by one additional point to the intersection of the PG$(4, 2)$ and $Q^{+}(7, 2)$. These double-five of points form five concurring lines, the point of concurrence being nothing but the additional point of the solid, and hence a point of $Q^{+}(7, 2)$, defined by the quartet of points of ${\cal O}$ that complements the pentad in question; a diagrammatical illustration of this feature, for ${\cal O^{*}}$, is supplied in Figure 7.

\begin{figure}[t]
\centerline{\includegraphics[width=7cm,clip=]{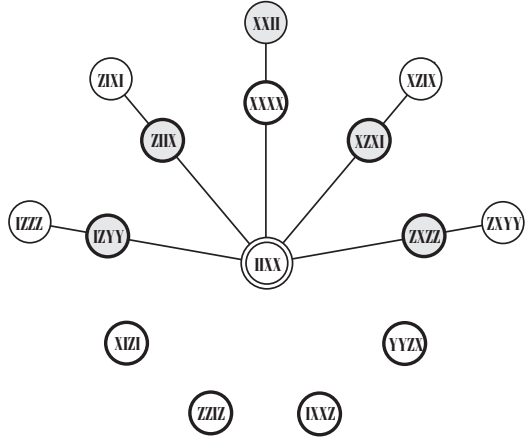}}
\vspace*{.2cm}
\caption{A sketchy outline of the intersection of PG$(4, 2)$ generated by a selected quintuple of points of  ${\cal O^{*}}$ (the five points in the upper half of the inner circle) and the quadric $Q^{+}(7, 2)$; highlighted by shading is one quartet of points within the pentad and the associated additional point of the solid defined by this quartet.}
\end{figure}

The attentive reader may have noticed that whereas for $s = 2, 3$ and $4$ subspaces PG$(s - 1, 2)$ defined by $s$-point subsets of ${\cal O}$ cut the quadric $Q^{+}(7, 2)$ in non-singular quadrics (of type $Q^{+}(1, 2)$, $Q(2, 2)$ and $Q^{-}(3, 2)$, respectively), the last discussed case, $s = 5$, falls short in this respect (for a set of concurrent lines does not represent any non-singular quadric). This is, however, the sole exception, since in the remaining two cases, $s=6$ and $s=7$, we again encounter non-singular quadrics, these being, respectively, of $Q^{-}(5, 2)$- and $Q(6, 2)$-types. We shall discuss in some detail only the former case, making just a brief comment on the latter one.

To understand how a $Q^{-}(5, 2)$ emerges as the intersection of the $Q^{+}(7, 2)$ and  PG$(5, 2)$ generated by a sextet of points of ${\cal O}$, it suffices to recall a well-known representation of  $Q^{-}(5, 2)$ as a generalized quadrangle GQ$(2, 4)$ where its 27 points are split into a set of 15 elements, forming a generalized quadrangle GQ$(2, 2)$, and a set called Schl\" afli double-six \cite{payt}. The split in the latter set is such that each of the six points in either set form with five points of the other set five lines lying fully on $Q^{-}(5, 2)$; the 30 lines one gets this way form 15 pairs, where two lines in each pair are concurrent and the 15 points of concurrence are nothing but the 15 points of the former set (GQ$(2, 2)$). Let us illustrate this in more detail on ${\cal O^{*}}$, as displayed in Figure 8. Here, the double six is highlighted by shading. One half of it comprises the selected sextet of points of ${\cal O^{*}}$, whereas its other half is represented by the six points of concurrence of the five on-quadric lines in each of ${6 \choose 5} = 6$ PG$(4, 2)$s defined by pentads of points within the sextet selected. (Otherwise rephrased, any such double-six arises as the symmetric difference of two ovoids having three points in common (see Figure 3).)  Moreover, six lines defined by pairs of associated points in the double six all pass through a common point, the nucleus of the conic defined by the triple of points of ${\cal O^{*}}$ that is the complement of the sextet.

\begin{figure}[t]
\centerline{\includegraphics[width=12.0cm,clip=]{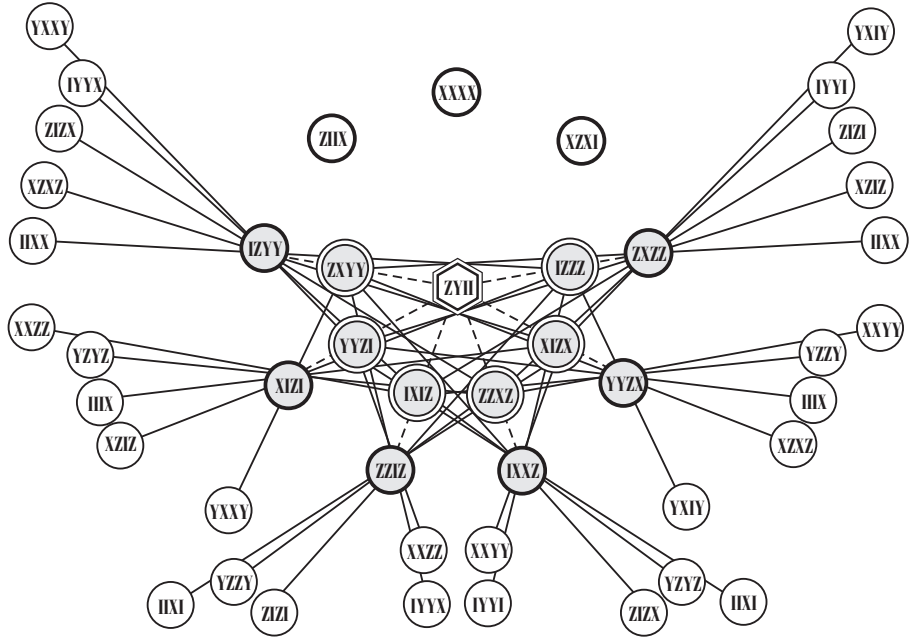}}
\vspace*{.2cm}
\caption{A schematic sketch illustrating intersection, $Q^{-}(5, 2)$, of the $Q^{+}(7, 2)$ and the subspace PG$(5, 2)$ spanned by a sextet of points (shaded) of ${\cal O^{*}}$; shown are all 27 points and 30 out of 45 lines of $Q^{-}(5, 2)$. Note that each point outside the double-six occurs twice; in the language of generalized quadrangles this corresponds to the fact that any two ovoids of GQ$(2, 2)$ have a point in common. The point $Z \otimes Y \otimes I \otimes I$ is the nucleus of the conic defined by three unshaded points of ${\cal O^{*}}$.}
\end{figure}

Finally, a short note on the intersection of the $Q^{+}(7, 2)$ and a PG$(6, 2)$ spanned by a heptad of points of ${\cal O}$. This intersection is isomorphic to a $Q(6, 2)$ whose nucleus is the third point of the line defined by the two points that are complementary to the heptad selected.

\subsection{Other Notable Subconfigurations of the Group}
In the last subsection we dealt with subsets/subconfigurations featuring mostly symmetric elements of $\overline{{\cal P}}_4$. This subsection will be focused on those ovoid-associated aggregates within $\overline{{\cal P}}_4$ in which the dominant role is played by skew-symmetric elements.

\begin{figure}[t]
\centerline{\includegraphics[width=9.0cm,clip=]{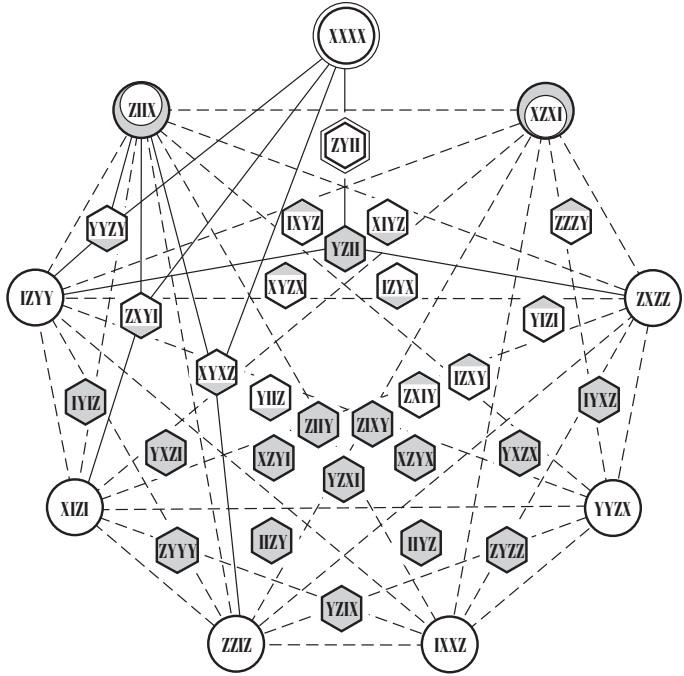}}
\vspace*{.2cm}
\caption{A set of nuclei (hexagons) of the 28 conics of ${\cal O^{*}}$ having a common point (double-circle); when one nucleus (double-hexagon) is discarded, the set of remaining 27 elements is subject to a natural $15 + 2 \times 6$ partition (illustrated by different types of shading).}
\end{figure}

We shall start with a configuration of 27 skew-symmetric elements that exhibits a natural $15 + 2 \times 6$ partition, similar to the above-described one within the point-set of $Q^{-}(5, 2)$. We have already mentioned an obvious fact that there are 28 triangles, and so 28 conics, on any point of ${\cal O}$. The nuclei of these conics are all distinct and the set we are interested in is obtained if one of the nuclei is singled out. The conic whose nucleus was singled out becomes distinguished within the set; hence, also its two points different from the common-to-all-the-conics point are distinguished. As on either of them there are other six conics in the set, the corresponding nuclei will form a double-six. We again take ${\cal O^{*}}$ to illustrate this property, as furnished by Figure 9. Here, the 28 conics all share the point $X \otimes X \otimes X \otimes X$ (double-circle), the nucleus of the selected conic is $Z \otimes Y \otimes I \otimes I$ (double-hexagon) and its two additional points (shaded crescents) are $Z \otimes I \otimes I \otimes X$ and $X \otimes Z \otimes X \otimes I$; the nuclei of the other six conics passing via the former/latter point are represented by hexagons shaded in the lower/upper part. It is also easily checked that six lines defined by pairs of associated points of the double six are concurrent, the point of concurrence, $Y \otimes Z \otimes X \otimes X$ , being the third point on the line defined by the common point ($X \otimes X \otimes X \otimes X$) and the singled-out nucleus ($Z \otimes Y \otimes I \otimes I$).
Despite their similarity, this configuration is fundamentally different from $Q^{-}(5, 2)$. This is because the third point on any of 30 lines constructed from the double-six in the same way as described above for $Q^{-}(5, 2)$  corresponds to a {\it symmetric} element of the group and the 15 points of concurrence of these lines thus cannot be identified with the complement of the double-six, which features exclusively {\it skew-symmetric} elements (see Figure 9). Nevertheless, as the interested reader can readily verify, these 15 symmetric elements and the double-six of skew-symmetric ones do form a configuration isomorphic to  $Q^{-}(5, 2)$.

Nuclei of ovoid-associated conics will also play a prominent role in configurations to be addressed next. Let us consider a set of nuclei of seven conics of ${\cal O^{*}}$ having two points in common.
If, without loss of generality,  the two shared points are taken to be  $Z \otimes Z \otimes I \otimes Z$ and $I \otimes X \otimes X \otimes Z$, we get a particular set of seven nuclei (skew-symmetric elements) as shown in Figure 10, {\it left}. Straightforward calculations yield that the remaining point on {\it each} of ${7 \choose 2} = 21$ lines, defined by pairs of points of the heptad, corresponds to a {\it skew-symmetric} element as well --- see Figure 10, {\it right}. So, we arrive at a set of 28 points, lying off the $Q^{+}(7, 2)$,  in which there exists, remarkably, a subset of seven points such that 21 lines defined by this seven-point subset are contained fully in the set. But this is exactly (see Sec.\,2) the defining property of a {\it Conwell heptad} of PG$(5, 2)$ with respect to a hyperbolic quadric $Q^{+}(5, 2)$ --- a set of seven out of 28 points lying off $Q^{+}(5, 2)$ such that the line defined by any two of them is skew to $Q^{+}(5, 2)$. This analogy with Conwell heptads can further be strengthened by observing that all 35 distinct nuclei of the conics defined by triples of points of our heptad, as illustrated in Figure 11, stand for {\it symmetric} operators of the group --- this being the counterpart to 35 points situated on $Q^{+}(5, 2)$.

\begin{figure}[t]
\centerline{\includegraphics[width=6.5cm,clip=]{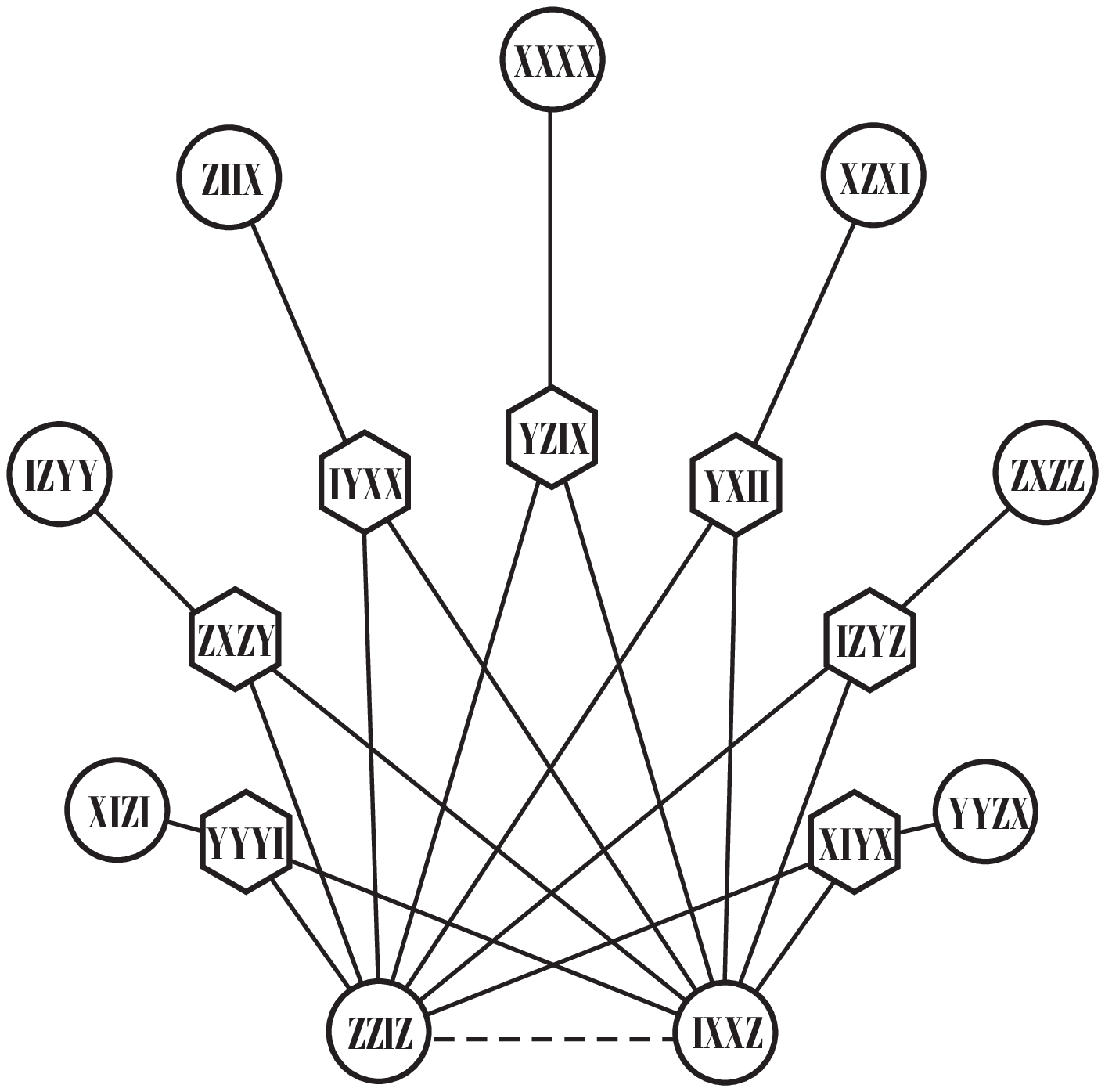}\hfill\includegraphics[width=6.5cm,clip=]{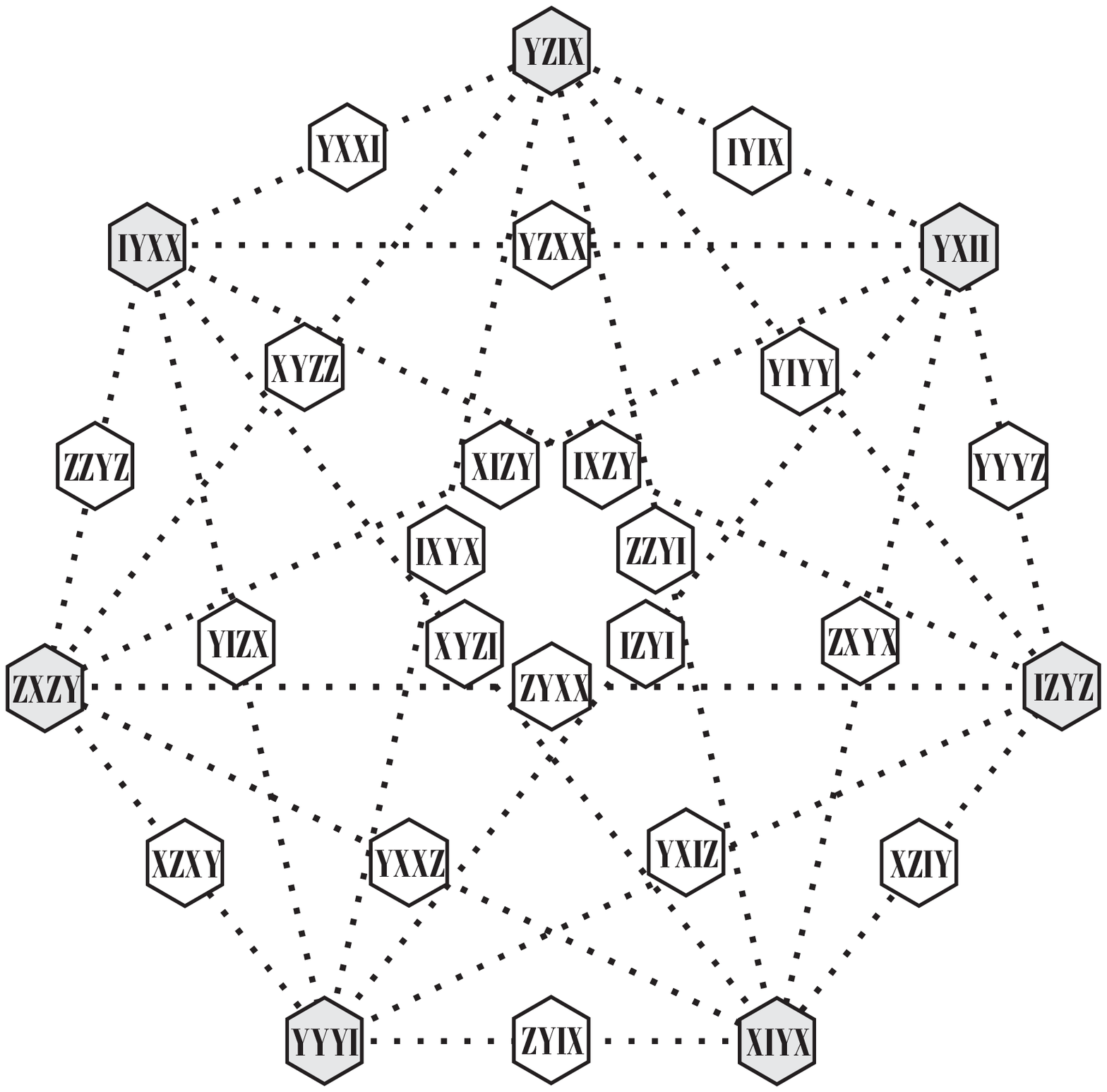}}
\vspace*{.2cm}
\caption{An illustration of the seven nuclei (hexagons) of the conics on two particular points of ${\cal O^{*}}$ ({\it left}) and the set of 21 lines (dashed) defined by these nuclei ({\it right}).}
\end{figure}

\begin{figure}[h]
\centerline{\includegraphics[width=7.1cm,clip=]{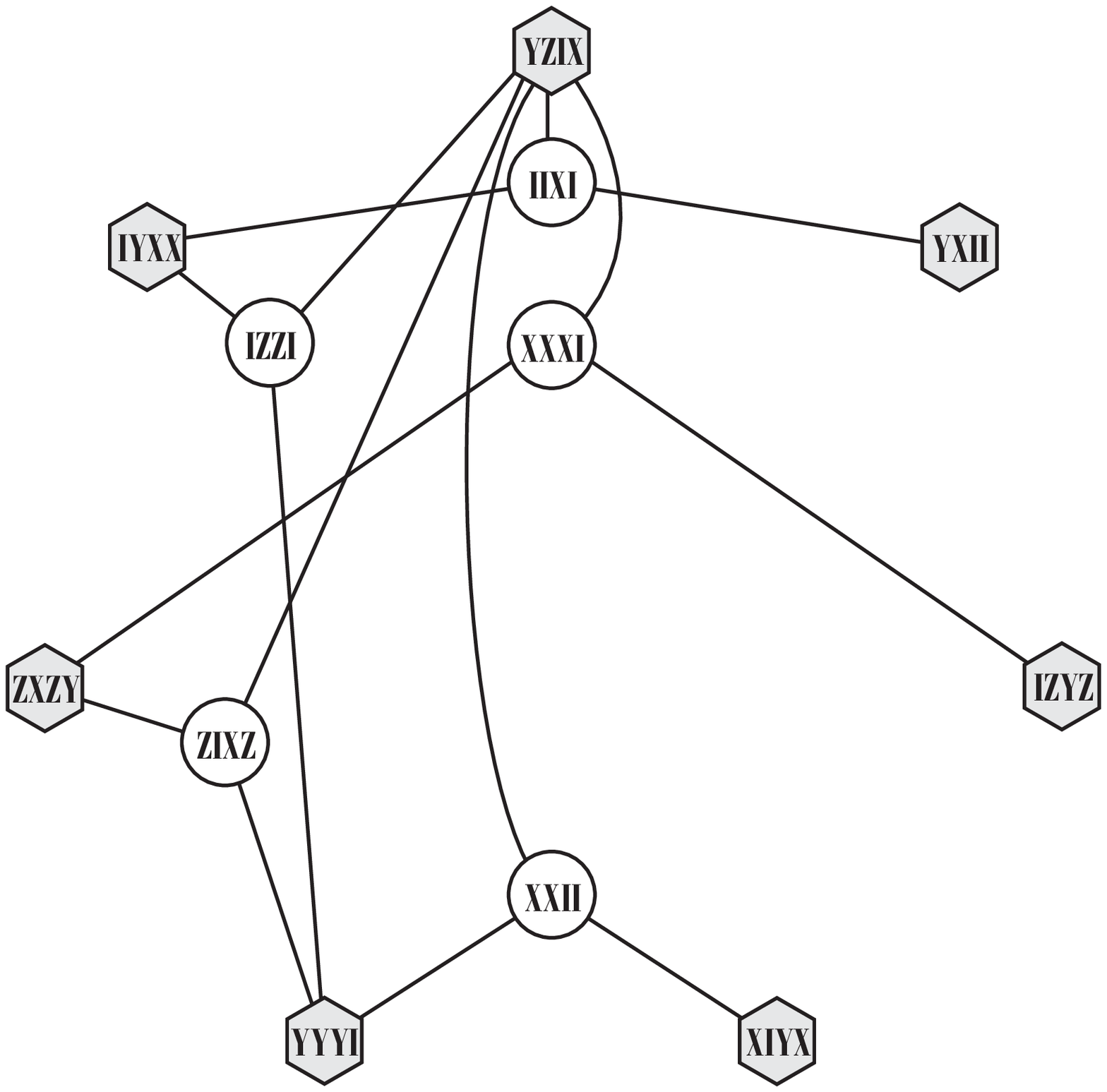}\hfill\includegraphics[width=7.1cm,clip=]{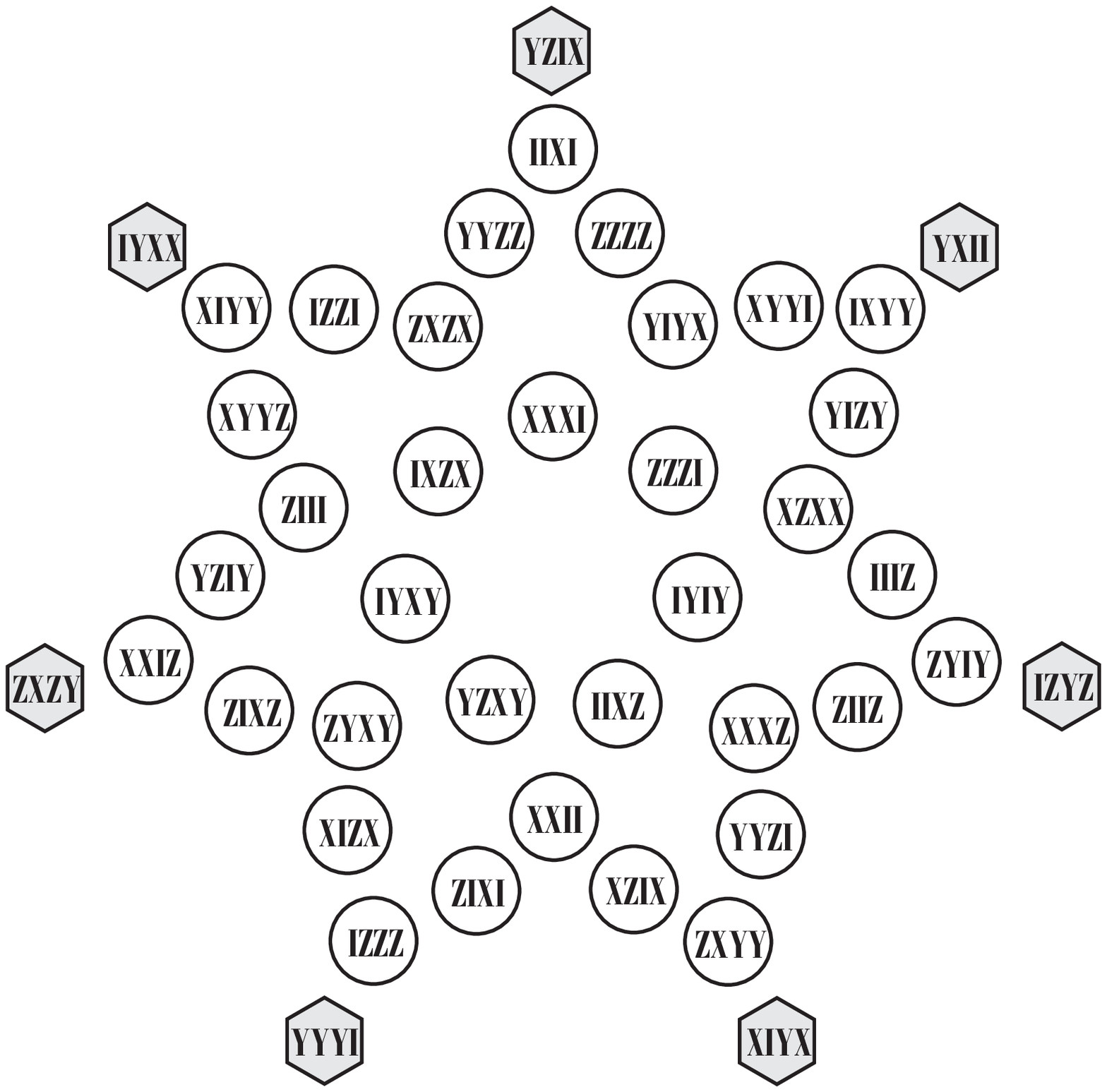}}
\vspace*{.2cm}
\caption{An illustration of the set of nuclei of the conics spanned by triples of points of our heptad, in a form of five different orbits under an automorphism of order seven.  The left-hand-side figure depicts just a single representative from each orbit, whilst the right-hand-side one shows all 35 points.}
\end{figure}

There exist several notable configurations of `Conwell heptads' associated with any given ovoid of $Q^{+}(7, 2)$. Perhaps the most pronounced of them, and most symmetric as well, is the one featuring three heptads sharing a point. This configuration is obtained if three pairs of common points of the conics form a triangle. Let us recall that our triangle lies in one more ovoid of  $Q^{+}(7, 2)$ (Figure 3). Hence, there is an additional set of three heptads on the same point.
Another very attractive configuration entails four consecutive `Conwell heptads' in the sense that the pairs of common points form a quadrangle. In such a configuration, a given heptad shares one point with either of the neighbouring heptads and is disjoint from the remaining heptad. The four shared points and the four vertices of the quadrangle can be paired in such a  way that the corresponding four lines are concurrent, their meet being the fifth point  of the solid spanned by the vertices of the quadrangle that is on $Q^{+}(7, 2)$.

\section{Conclusion}
We have carried out a comprehensive examination of the geometrical structure of the real four-qubit Pauli group in terms of ovoids lying on the distinguished hyperbolic quadric $Q^{+}(7, 2)$ that is the locus of all symmetric elements of the group. We have found, and described in detail, a number of remarkable subconfigurations of the group that can be divided into two groups: those related to the intersections of the $Q^{+}(7, 2)$ with projective subspaces spanned by various subsets of points of an ovoid and those comprising various aggregates of the nuclei of conics defined by an ovoid. About a dozen of distinguished types of configurations have been found, each being diagrammatically illustrated on the same, particularly-chosen ovoid.  Amongst the most interesting one can rank: a subset of 12 skew-symmetric elements lying on four mutually skew lines that span the whole ambient space,  a subset of 15 symmetric elements that corresponds to two ovoids having a triple of points in common, a subset of 19 symmetric elements generated by two ovoids on a common point, a subset of 27 symmetric elements that can be partitioned into three ovoids in two unique ways,  a subset of 27 skew-symmetric elements that exhibits a $15 + 2 \times 6$ split reminding that found on a $Q^{-}(5, 2)$, and a subset of seven skew-symmetric elements that is an analogue of a Conwell heptad of PG$(5, 2)$. The strategy  we employed is completely novel and unique in its nature, as are the results obtained.

As already stressed, generalized Pauli groups are well known to physicists and play a very important role in quantum information theory. It is, therefore, desirable to deepen our understanding of their geometrical structure. The present paper, in our opinion, represents a substantial contribution in this respect as per the four-qubit case. One of the reasons  why we focused on  this particular  case is our belief that it(s symplectic geometry) may shed some light on the mystery of the so-called black-hole-qubit analogy/correspondence (for a relatively recent review, see \cite{bor}).
Another reason is that hyperbolic quadrics $Q^{+}(2N - 1, 2)$ for  $N \geq 5$ have no ovoids \cite{kan}. Since both  $Q^{+}(3, 2)$ and $Q^{+}(5, 2)$ do feature ovoids, the $N=4$ case is the last one in the hierarchy enjoying this property; this fact is also telling us about the prominence of two-, three- and four-qubit Pauli groups.

\section*{Acknowledgement}
This work was partially supported by the Slovak grant agency VEGA, project No. 2/0098/10, and the New Hungary Development Plan T\' AMOP-4.2.1/B-09/1/KMR-2010-002. We are grateful to Dr. Richard Kom\v z\'{\i}k for computer assistance.


\normalsize
\vspace*{-.1cm}
\begin{thebibliography}{10}
\itemsep=-2pt
\bibitem{sp}
M. Saniga and M. Planat, Adv. Studies Theor. Phys. 1 (2007) 1--4; arXiv:quant-ph/0612179.
\bibitem{ps}
M. Planat and M. Saniga, Quant. Inform. Comput. 8 (2008) 127--146; arXiv:quant-ph/0701211.
\bibitem{spp}
M. Saniga, M. Planat and P. Pracna, Theor. Math. Phys. 155 (2008)  905--913; arXiv:quant-ph/0611063.
\bibitem{hos}
H. Havlicek, B. Odehnal and M. Saniga, SIGMA 5 (2009) Art. No. 96; arXiv:0903.5418.
\bibitem{th}
K. Thas, EPL / Europhysics Letters 86 (2009) Art. No. 60005.
\bibitem{pla}
M. Planat, J. Phys. A: Math. Theor. 44 (2011) Art. No. 045301.
\bibitem{ks}
S. Kochen and E. Specker, J. Math. Mechanics 17 (1967) 59--87.
\bibitem{lsv}
P. L\' evay, M. Saniga and P. Vrana, Phys. Rev. D78 (2008) Art. No. 124022; arXiv:0808.3849.
\bibitem{vl}
P. Vrana and P. L\' evay, J. Phys. A: Math. Theor. 43 (2010) Art. No. 125303; arXiv:0906.3655.
\bibitem{sl}
M. Saniga and P. L\' evay, EPL / Europhysics Letters 97 (2012) Art. No. 50006; arXiv:1111.5923.
\bibitem{study}
E. Study, Nachrichten K. Gesellschaft Wiss. G\"ottingen (Math.-phys. Klasse) 1912, 433--479.
\bibitem{tits}
J. Tits, Inst. Hautes Etudes Sci. Publ. Math. 2 (1959)
13--60.
\bibitem{edge1}
W. L. Edge, Ann. Mat. Pura Appl. 61 (1963) 1--95.
\bibitem{lsvp}
P. L\' evay, M. Saniga, P. Vrana and P. Pracna, Phys. Rev. D79 (2009) Art. No. 084036; arXiv:0903.0541.
\bibitem{ht}
J. W. P. Hirschfeld and J. A. Thas,  General Galois Geometries, Oxford University Press, Oxford, 1991.
\bibitem{cam}
P. J. Cameron, Projective and Polar Spaces, a manuscript available on-line from http://www.maths.qmw.ac.uk/$\widetilde{~~}$pjc/pps/.
\bibitem{thas}
J. A. Thas, J. Combin. Theory A56 (1991) 303--308.
\bibitem{con}
G. M. Conwell, Ann. Math. 11 (1910) 60--76.
\bibitem{nc}
M. Nielsen and I. Chuang, Quantum Computation and Quantum Information,
Cambridge University Press, Cambridge, 2000.
\bibitem{cald}
A. R. Calderbank, E. M. Rains, P. W. Shor and N. J. A. Sloane, Phys. Rev. Lett. 78 (1997) 405--408; arXiv:quant-ph/9605005.
\bibitem{payt}
S. E. Payne and J. A. Thas, Finite Generalized Quadrangles, EMS Series of Lectures in
Mathematics, European Mathematical Society, Z\" urich, 2009.
\bibitem{bor}
L. Borsten, D. Dahanayake, M. J. Duff, H. Ebrahim, and W. Rubens, Phys. Reports 471 (2009) 113--219; arXiv:0809.4685.
\bibitem{kan}
W. M. Kantor, Ann. Discrete Math. 18 (1983) 511--518.
\end{thebibliography}
\end{document}